\documentclass[epjCONF, columns]{svjour} 
\usepackage{graphics}
\usepackage[varg]{txfonts} 
\usepackage[latin1]{inputenc}
\session-title{2011 Hadron Collider Physics symposium}
\begin{document}
%
\title{ATLAS Exotic Searches}
\author{Nicolas Bousson, On behalf of the ATLAS Collaboration\thanks{\email{nicolas.bousson@cern.ch}}}
\institute{Centre de Physique des Particules de Marseille, IN2P3-CNRS/Aix-Marseille-Universit\'e}
\abstract{ Thanks to the outstanding performance of the Large Hadron Collider (LHC) that delivered more than 2~fb$^{-1}$ of proton-proton collision data at center-of-mass energy of 7 TeV, the ATLAS experiment has been able to explore a wide range of exotic models trying to address the questions unanswered by the Standard Model of particle physics. Searches for leptoquarks, new heavy quarks, vector-like quarks, black holes, hidden valley and contact interactions are reviewed in these proceedings.
} 
\maketitle
\section{Introduction}
\label{intro}
While the observation of the Higgs boson would explain the electro-weak (EW) symmetry breaking in the Standard Model (SM), it still leaves many unexplained fundamental problems such as the hierarchy problem, the fermion mass hierarchy, dark matter, the baryon asymmetry of the Universe, etc. Many models have been proposed over the past decades which address the issues above and motivate a wide range of searches by the ATLAS experiment. Searches for new particles, strong gravity, and more generic searches for new phenomena with over 1~fb$^{-1}$ of LHC pro-ton-proton collision data at center-of-mass energy of 7 TeV collected by the ATLAS experiment are presented in the following.
\section{Searches for new particles}
\label{sec:1}
\subsection{Leptoquarks}
Leptoquarks (LQ) are colour-triplet bosons that carry both lepton and baryon numbers, and fractional electric charge. They are introduced by various extension of the SM, such as Grand Unification models. A search for pair-produced LQs assumed to couple only to quarks and leptons of the first SM generation is described in Ref.~\cite{LQ}, using 1.03 fb$^{-1}$ of data, for two different final states. In the first one both LQs decay into an electron and a quark, while in the second final state one of the LQs decays into an electron and a quark and the other LQ decays into an electron-neutrino and a quark. These result in two different experimental signatures. One such signature is the production of two electrons and two jets and the other one comprises one electron, two jets, and missing transverse momentum. 

\begin{figure}
\begin{center} 
\resizebox{1.0\columnwidth}{!}{
  \includegraphics{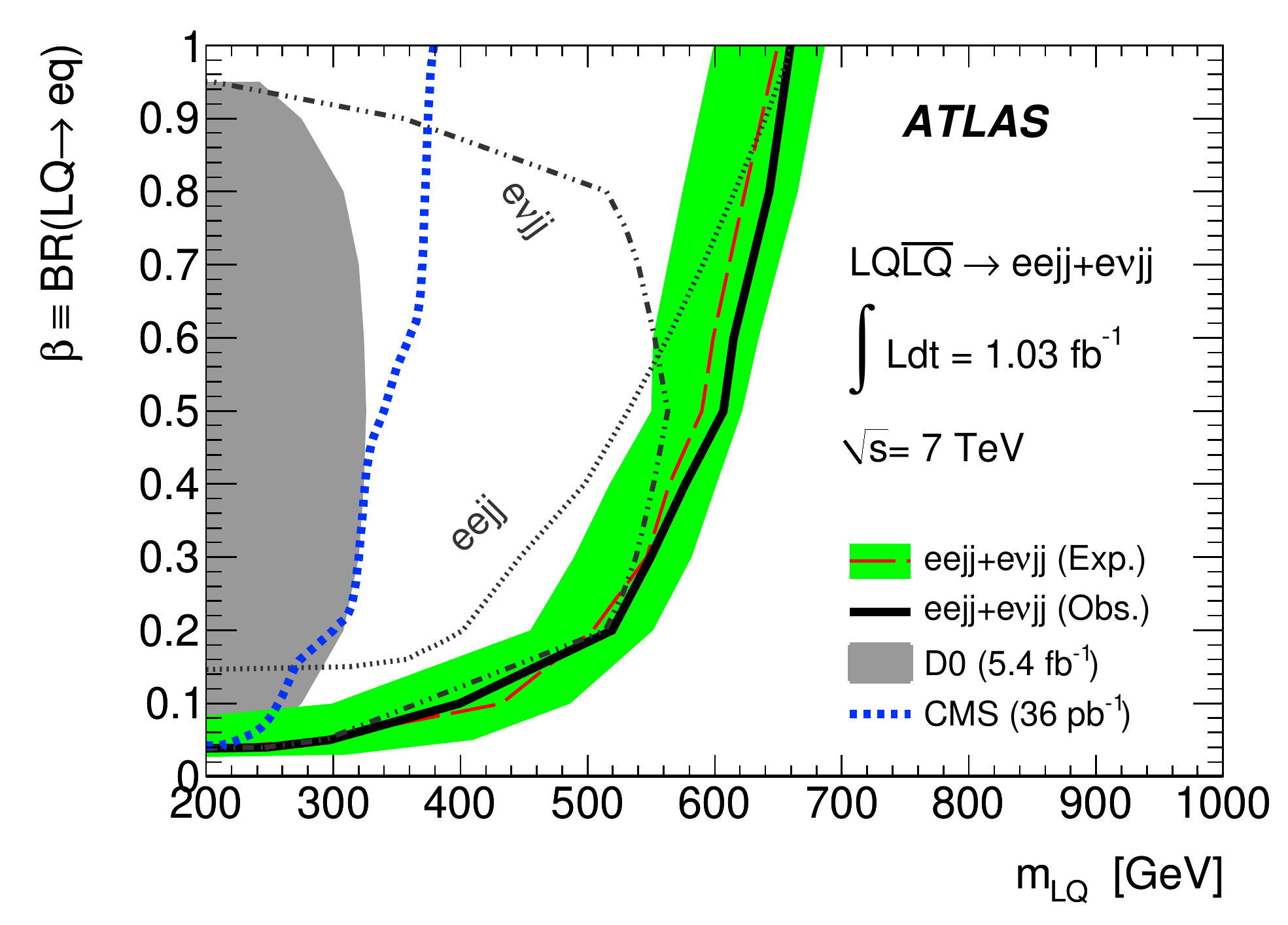} }
\caption{95\% CL LQ exclusion regions shown in the $\beta$ versus LQ mass plane. 
The solid band contains 68\% of possible outcomes
from pseudo-experiments.}
\label{fig:lepto}       
\end{center}
\end{figure}

No excess over SM background expectations is obser-ved in the data, and 95\% confidence level (CL) upper bou-nds on the production cross-section are thus determined. The results from the two final states are combined and presented in the $\beta$ versus LQ mass plane in Fig.~\ref{fig:lepto}, where $\beta$ is the branching ratio for a single LQ to decay into a charged lepton and a quark. For $\beta=1$, LQ masses up to 660 GeV are excluded. 

\subsection{New heavy quarks}
New heavy quarks are among the simplest extensions to the SM. A search for pair-produced exotic top partners $T\bar{T}$, each decaying to a top quark and a stable, neutral weakly-interacting particle $A_0$, has been conducted in Ref.~\cite{TT} with 1.04 fb$^{-1}$ of data. The search is performed in the $t\bar{t}$ single-lepton channel where one $W$ boson produced by the top pair decays to a lepton-neutrino pair and the other $W$ boson decays to a pair of quarks, resulting in a final state with an isolated lepton, four or more jets and large missing transverse momentum. Assuming a $T\bar{T} \rightarrow t\bar{t}A_0A_0$ branching ratio of 100\%, $T$ masses up to 420 GeV are excluded at the 95\% CL for an $A_0$ mass below 10 GeV. This limit is also valid for different models that have a final state identical to the one presented here. In particular, it can be applied to the case of a fourth sequential SM generation, that could have major implications, $e.g.$ in cosmology. 

\subsection{Vector-like quarks}
\label{sec:vlq}

\begin{figure}
\begin{center} 
\resizebox{0.6\columnwidth}{!}{
  \includegraphics{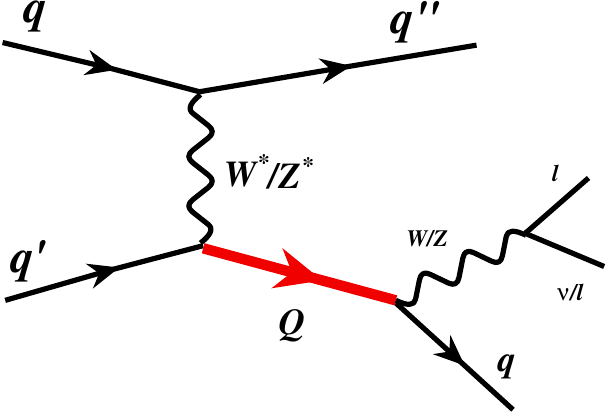} }
\caption{Vector-like quark production and decay diagram for the t-channel process (charged current or neutral current). The thick line indicates the vector-like quark.}
\label{fig:vlqfeynman}       
\end{center}
\end{figure}

While a fourth generation of chiral fermions may exist, there are actually strong constraints on it from EW measurements. Vector-like (VL) fermions escape these constraints, and are predicted by many extensions of the SM. They would be a novel form of matter. In certain scenarios, $e.g.$ extra-dimensions, VL quarks $Q$ could couple sizeably to the light generation of quarks, leading to a strong signal at the LHC. For single production of such quarks, the $t$-channel, illustrated in Fig.~\ref{fig:vlqfeynman}, is dominant. A search for experimental signatures with two jets and, either two leptons, or one lepton and missing transverse momentum, is described in Ref.~\cite{VLQ}, using 1.04 fb$^{-1}$ of data. 

\begin{figure}
\begin{center} 
\resizebox{1.0\columnwidth}{!}{
  \includegraphics{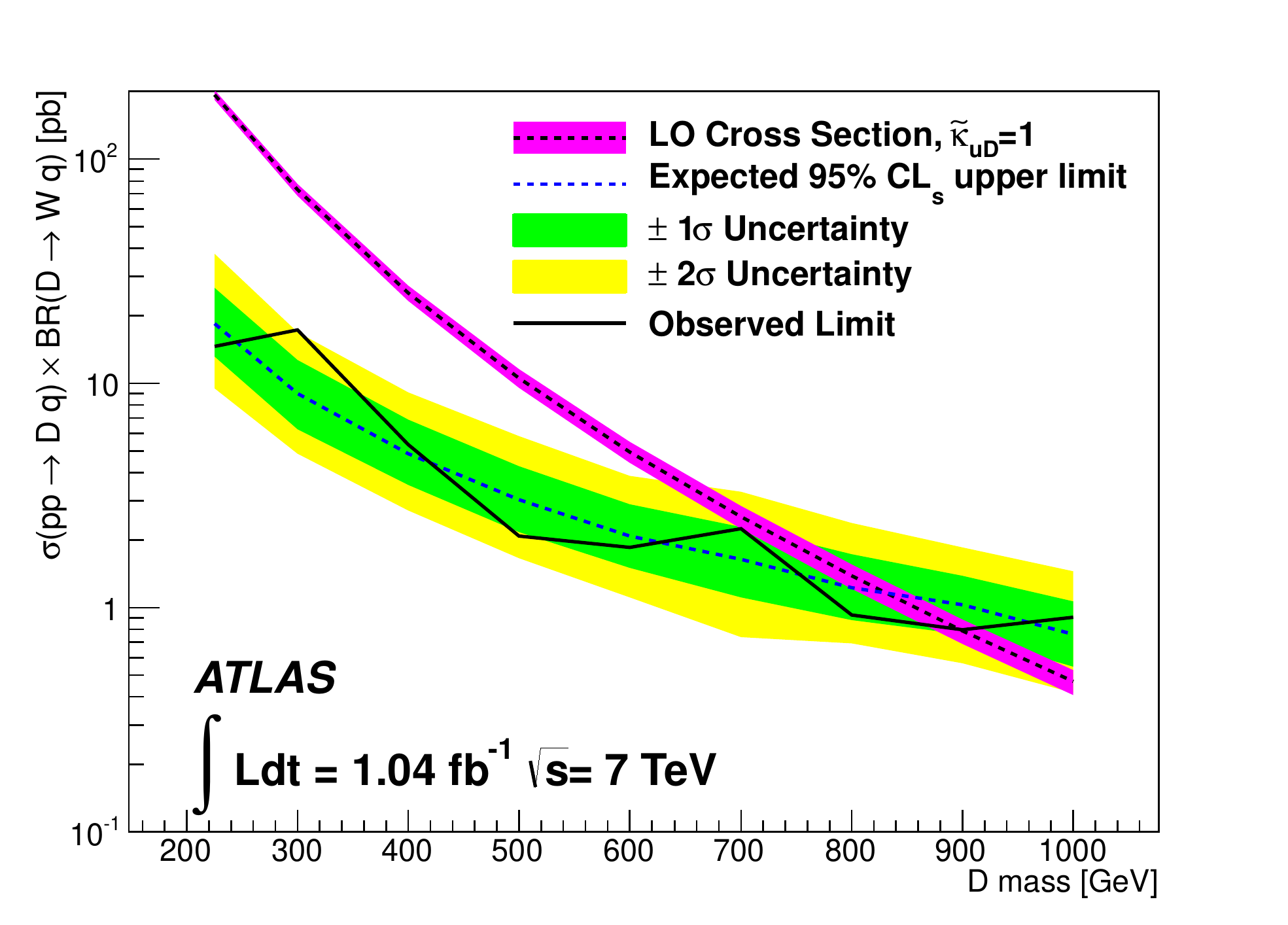} }
\caption{Upper limits at 95\% CL on the cross-section times branching ratio $\sigma(pp \rightarrow Qq) \times BR(Q \rightarrow Vq)$ for the charged current channel as a function of mass.}
\label{fig:vlq}       
\end{center}
\end{figure}

No evidence is found for such quarks, and upper limits on the production cross-section times branching ratio to a vector boson and a quark are determined assuming reduced couplings of 1 at the $qVQ$ vertex ($V=W$ or $Z$) in the diagram in Fig.~\ref{fig:vlqfeynman}, as described in~\cite{VLQ}. VL quarks masses are excluded up to 900 GeV at the 95\% CL for the charged current channel, as shown in Fig.~\ref{fig:vlq}, and up to 760 GeV for the neutral current channel. These limits, which can be used to constrain different models of VL quarks~\cite{VLQ2}, are the most stringent to date on this model.

\section{Searches for strong gravity}
Models introducing extra dimensions can provide a solution to the hierarchy problem, which is that the Planck scale $M_{Pl}\sim 10^{16}$ TeV is higher than the EW scale $M_{EW}$ by many orders of magnitude. In these models, the Planck scale in (n+4) dimensions, $M_D$, would be much smaller than in four dimensions, because gravity propagates in all dimensions. If $M_D$ is in the TeV range, microscopic black holes (BH) could appear at the LHC, and evaporate by Hawking radiation. While uncertainties on these models are large due to our ignorance of quantum gravity, classical approximation for BH production and semi-classical approximation for decay are assumed valid for BH masses higher than a mass threshold, $M_{TH} \gg M_D$.

\subsection{Search with same-sign dimuon events}
\begin{figure}
\begin{center} 
\resizebox{1.0\columnwidth}{!}{
  \includegraphics{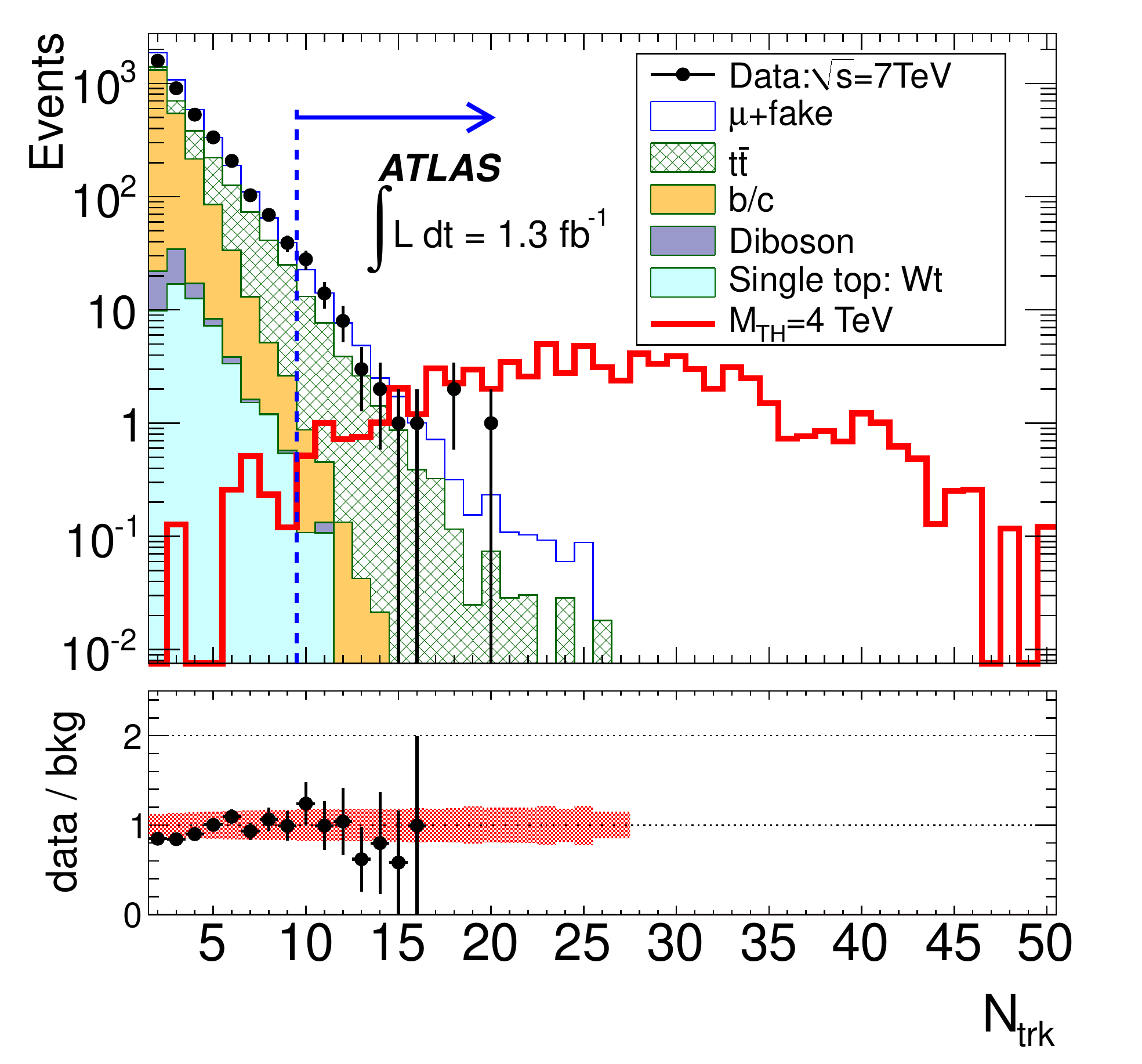} }
\caption{The track multiplicity distribution for same-sign dimuon events. The background histograms are stacked. The signal expectation for a non-rotating black hole model with parameters $M_D$ = 800 GeV, $M_{TH}$ = 4 TeV, and six extra dimensions is overlaid for illustrative purposes.}
\label{fig:strong1}       
\end{center}
\end{figure}

 As multiplicities of emitted particles are determined by the number of degrees of freedom of each particle type and their decay modes, BH events should have high multiplicity of tracks with high transverse momentum $p_T$. This is exploited in Ref.~\cite{BH1}, using 1.3 fb$^{-1}$ of data. Figure~\ref{fig:strong1} shows the number of tracks with $p_T>10$ GeV for events containing two muons with the same charge. This latter requirement allows to lower SM backgrounds while retaining a good signal acceptance. The region with $N_{trk} \ge 10$ is selected as the signal region. 

\begin{figure}
\begin{center} 
\resizebox{1.0\columnwidth}{!}{
  \includegraphics{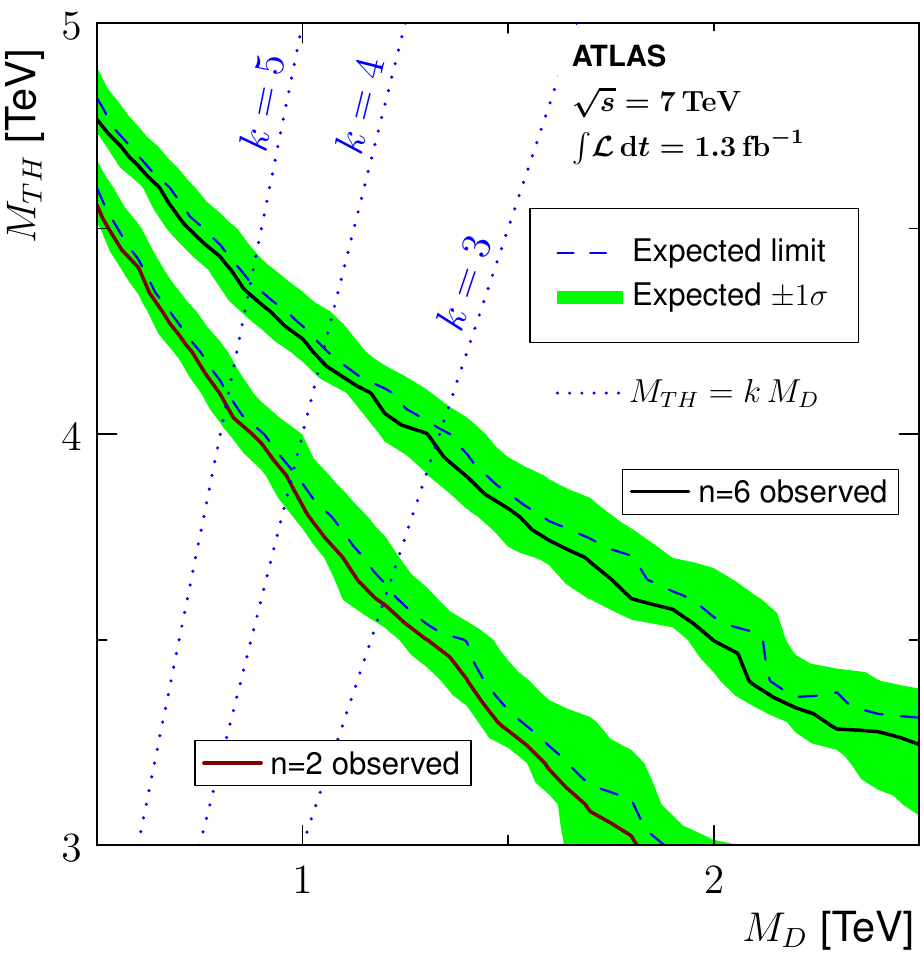} }
\caption{95\% CL exclusion contours for non-rotating black holes in models with two and six extra dimensions. The solid lines show the observed exclusion contour. Regions below the contour are excluded. }
\label{fig:bh}       
\end{center}
\end{figure}

No excess over the SM predictions is observed in the data. Figure~\ref{fig:bh} shows exclusion regions that are derived in the $M_{TH}$ versus $M_D$ plane. For $k=M_{TH}/M_D=5$, which is a physically favored region, and $n=6$, $M_{TH}$ up to $\sim 4.3$ TeV are ruled out. 

\subsection{Search with lepton+jets events}
Another way to exploit the high multiplicity of emitted particles is to search for excess of multi-object events produced at high $\Sigma p_T$ of leptons and jets. A search is conducted in Ref.~\cite{BH2} selecting events with one isolated lepton with $p_T > 40$ GeV, at least three reconstructed objects with $p_T>100$ GeV, at least one of which being a lepton, and $\Sigma p_T>1500$ GeV. 

\begin{figure}
\begin{center} 
\resizebox{1.0\columnwidth}{!}{
  \includegraphics{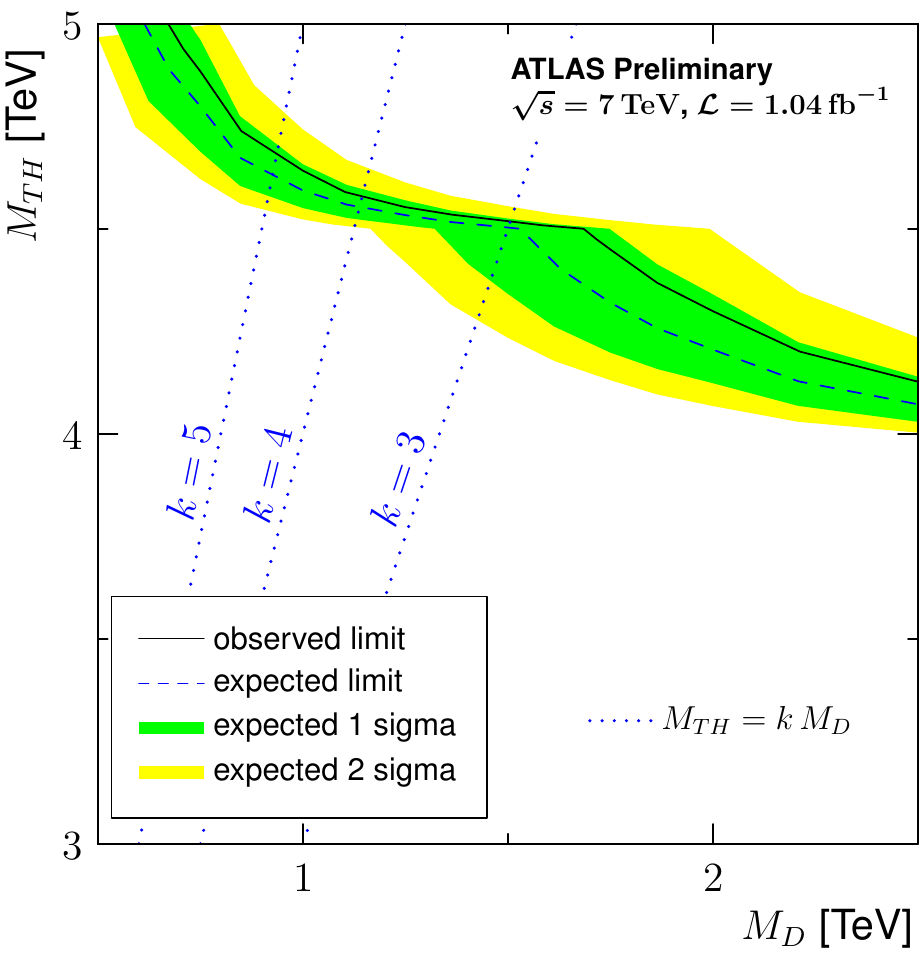} }
\caption{95\% CL limit in the $M_{TH} - M_D$ plane, for a rotating black hole model with six extra dimensions and a decay ending in a high-multiplicity remnant state. }
\label{fig:strong2}       
\end{center}
\end{figure}

No excess over the SM predictions is observed in 1.04 fb$^{-1}$ of data, and exclusion regions are derived, which can be seen on Fig.~\ref{fig:strong2} for a rotating BH. For $k=5$, $M_{TH}$ up to $\sim 4.7$ TeV are ruled out. 

\section{Generic searches for new phenomena}

\subsection{Hidden valley}
\begin{figure}
\begin{center} 
\resizebox{1.\columnwidth}{!}{
  \includegraphics{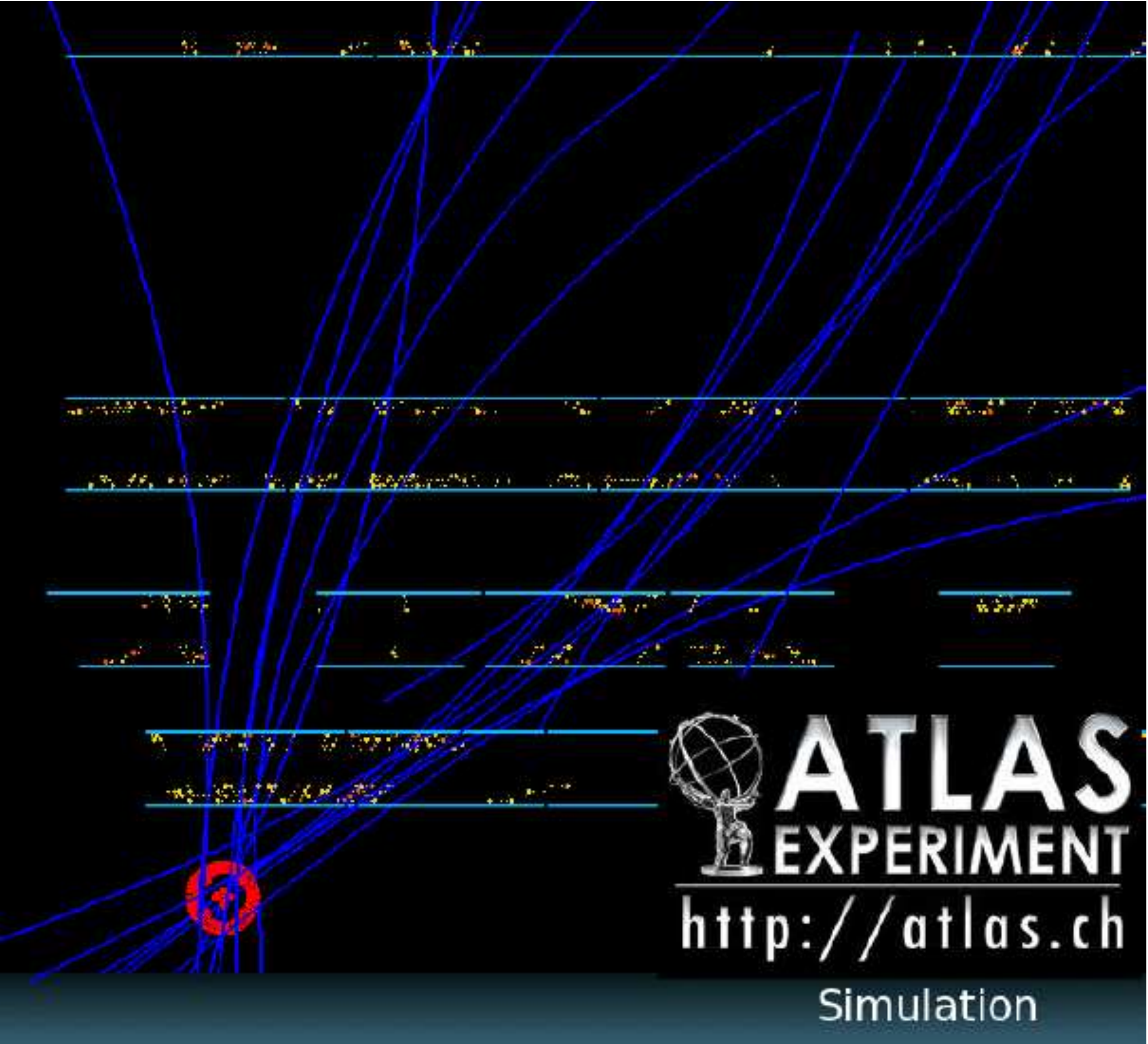} }
\caption{Event display for a simulated signal event showing reconstructed tracks and vertex position for a $\pi_v$ decaying just beyond the hadronic calorimeter. The reconstructed vertex position is shown by the red circle.}
\label{fig:hv2}       
\end{center}
\end{figure}

Many extensions of the SM include a light Higgs boson ($h^0$), with a mass below 140 GeV, that decays to neutral, weakly-coupled particles that can be long-lived. For example, in Hidden Valley (HV) scenarios, a new hidden sector is weakly coupled to the SM through a communicator that interacts with both sectors.
A search for the Higgs decay, $h^0 \rightarrow \pi_v \pi_v$, where the $\pi_v$ is a weakly-interacting pseudoscalar and has a displaced decay to fermion anti-fermion pairs $f\bar{f}$ has been conducted
 with 1.94 fb$^{-1}$ of data.
Due to the helicity suppression of the low-mass $f\bar{f}$ pairs, the $\pi_v$ decays predominantly to heavy fermions ($b\bar{b}$, $c\bar{c}$ and $\tau^+ \tau^-$) in the ratio 85:5:8 \%. However, the lifetime of the neutral states is not specified and can have a very large range. This analysis focuses on $\pi_v$ decays occuring in the Muon Spectrometer (MS) of the ATLAS detector. To improve background rejection, both $\pi_v$'s are required to decay in the MS. The analysis uses specialized tracking and vertexing reconstruction algorithms to reconstruct the vertices in the MS, from $\pi_v$ decays to jet pairs. This is illustrated in Fig.~\ref{fig:hv2}. For the event selection, a dedicated, signature-driven trigger, was developed to trigger on events with a $\pi_v$ decaying in the MS.

\begin{figure}
\begin{center} 
\resizebox{1.\columnwidth}{!}{
  \includegraphics{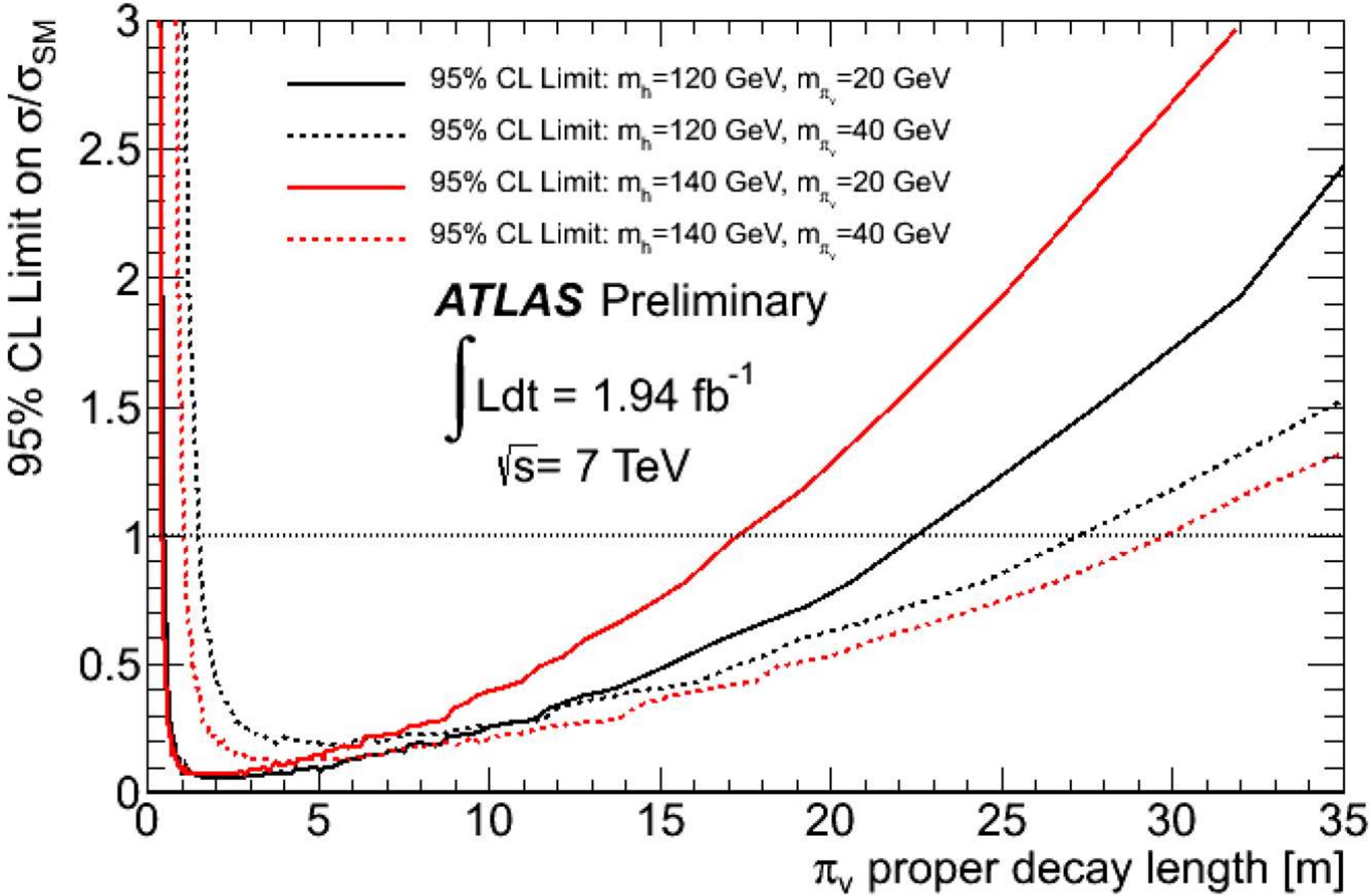} }
\caption{The observed 95\% CL upper limits on the process $h^0 \rightarrow \pi_v \pi_v$ expressed as the ratio to the SM cross-section for Higgs production. The exclusion limits assume 100\% branching ratio for the Higgs decaying to $\pi_v$s.}
\label{fig:hv}       
\end{center}
\end{figure}

No events are found in data which pass the selection requiring two isolated vertices in the MS. Figure~\ref{fig:hv} shows the upper limits on the cross-section for the process $h^0 \rightarrow \pi_v \pi_v$ normalised to the SM one that are derived at the 95\% CL as a function of the $\pi_v$ proper decay length. Table~\ref{tab:hvtable} shows the broad range of $\pi_v$ proper decay lengths which have been excluded for different $h^0$ and $\pi_v$ masses. These are the first limits on long-lived particles decaying to jet pairs.
\begin{table}
\begin{center} 
\caption{The excluded proper decay lengths $(c\tau )$, at 95\%
CL, of the $\pi_v$ for each of the signal samples. }
\resizebox{1.0\columnwidth}{!}{
  \includegraphics{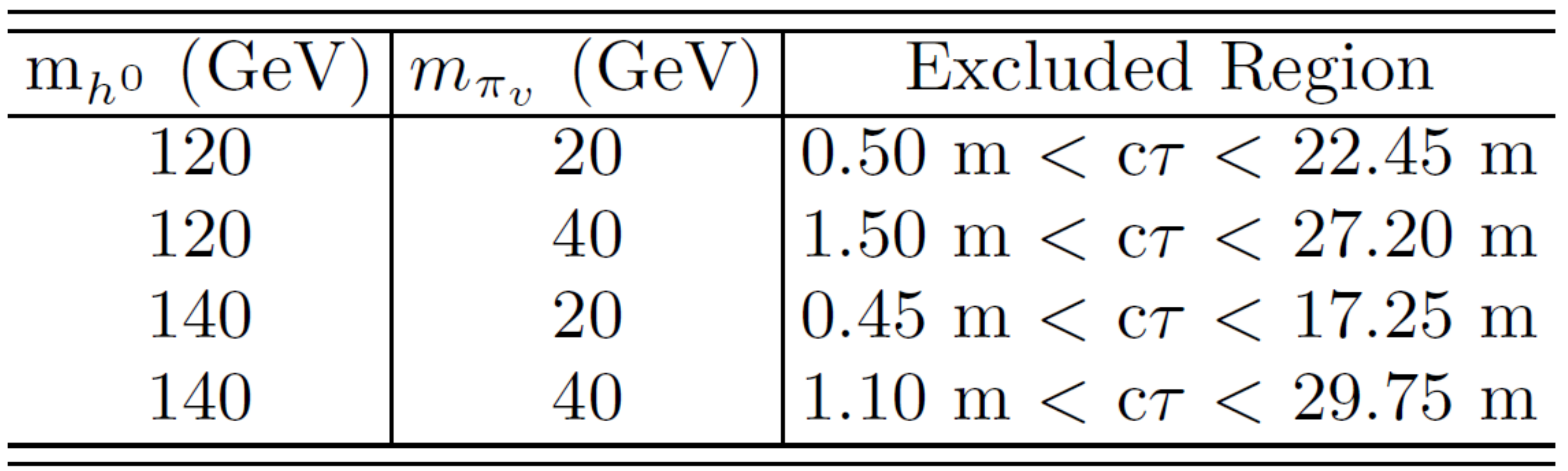} }
\label{tab:hvtable}       
\end{center}
\end{table}
\subsection{Contact interactions}
A wide range of new phenomena can produce modifications to the dilepton mass spectra predicted by the SM such as quark-lepton compositeness, extra dimensions, and new gauge bosons. The predicted form of these deviations is often either a resonance or an excess in the number of events in the spectra at high mass. If quarks and leptons are composite, with at least one common constituent, the interaction of these constituents would likely be manifested through an effective four-fermion contact interaction (CI) at energies well below the compositeness scale $\Lambda$. Such a CI could also describe a new interaction with a messenger too heavy for direct observation at the LHC.

In the context of the left-left isoscalar model, which is commonly used as a benchmark for CI searches, the differential cross-section for the process $q\bar{q} \rightarrow l^+ l^-$ can be written
\begin{equation}
\frac{d\sigma}{dm_{ll}} = \frac{d\sigma_{DY}}{dm_{ll}}-\eta_{LL}\frac{F_I(m_{ll})}{\Lambda^2}+\frac{F_C(m_{ll})}{\Lambda^4}
\end{equation}
where $m_{ll}$ is the final-state dilepton mass, and $\eta_{LL} = \pm 1$, a parameter defining the chiral structure of the new interaction. The expression above includes a SM Drell-Yan (DY) term, as well as DY-CI interference ($F_I$) and pure CI ($F_C$) terms.

\begin{figure}
\begin{center} 
\resizebox{1.0\columnwidth}{!}{
  \includegraphics{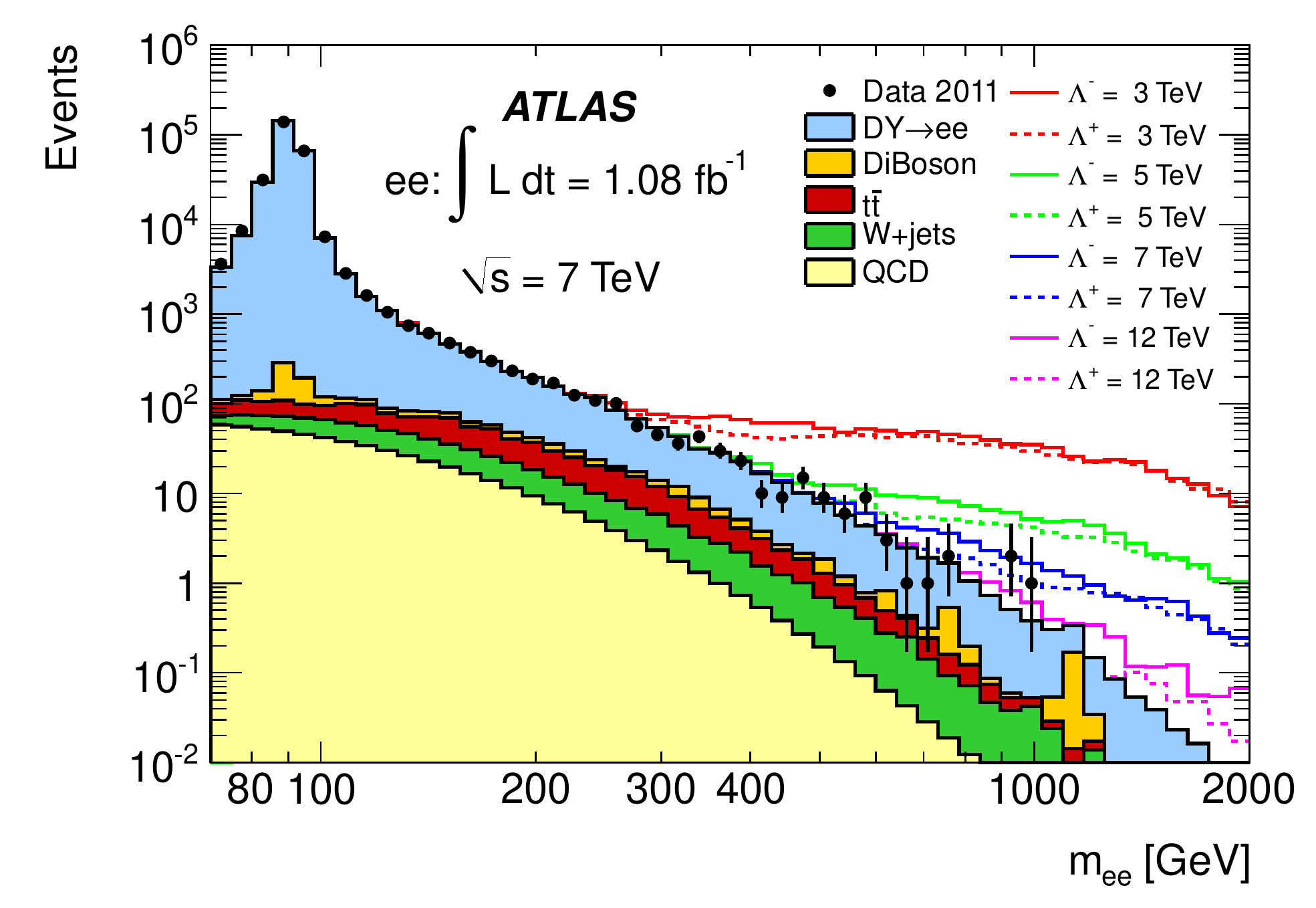} }
\caption{Dielectron invariant mass distribution for data (points) and Monte Carlo simulation (histograms). The open histograms correspond to the distributions expected in the presence of CIs with different values of $\Lambda$ for both constructive ($\Lambda^-$, $\eta_{LL} = -1$) and destructive ($\Lambda^+$, $\eta_{LL} = +1$) interference. }
\label{fig:ci}       
\end{center}
\end{figure}

A search for CI in dilepton channel is performed with 1.08 fb$^{-1}$ of data for the electron channel and 1.21 fb$^{-1}$ for the muon channel~\cite{CI}. Figure~\ref{fig:ci} shows the dilepton mass spectra in the electron channel in data, and in simulation for constructive and destructive interferences. There is no significant evidence for CI in the analysed data and thus limits have been set on the CI scale $\Lambda$ at 95\% CL, using two kinds of flat priors, which are summarized in Table~\ref{tab:citable}. These limits are the most stringent to date on $\mu\mu qq$ CIs and exceed the best existing limits set by a single experiment on $eeqq$ CIs for light-quark flavors.
\begin{table}
\begin{center} 
\caption{Expected and observed 95\% CL lower limits on the
CI energy scale $\Lambda$ for the electron and muon
channels, as well as for the combination of those channels.
}
\resizebox{1.0\columnwidth}{!}{
  \includegraphics{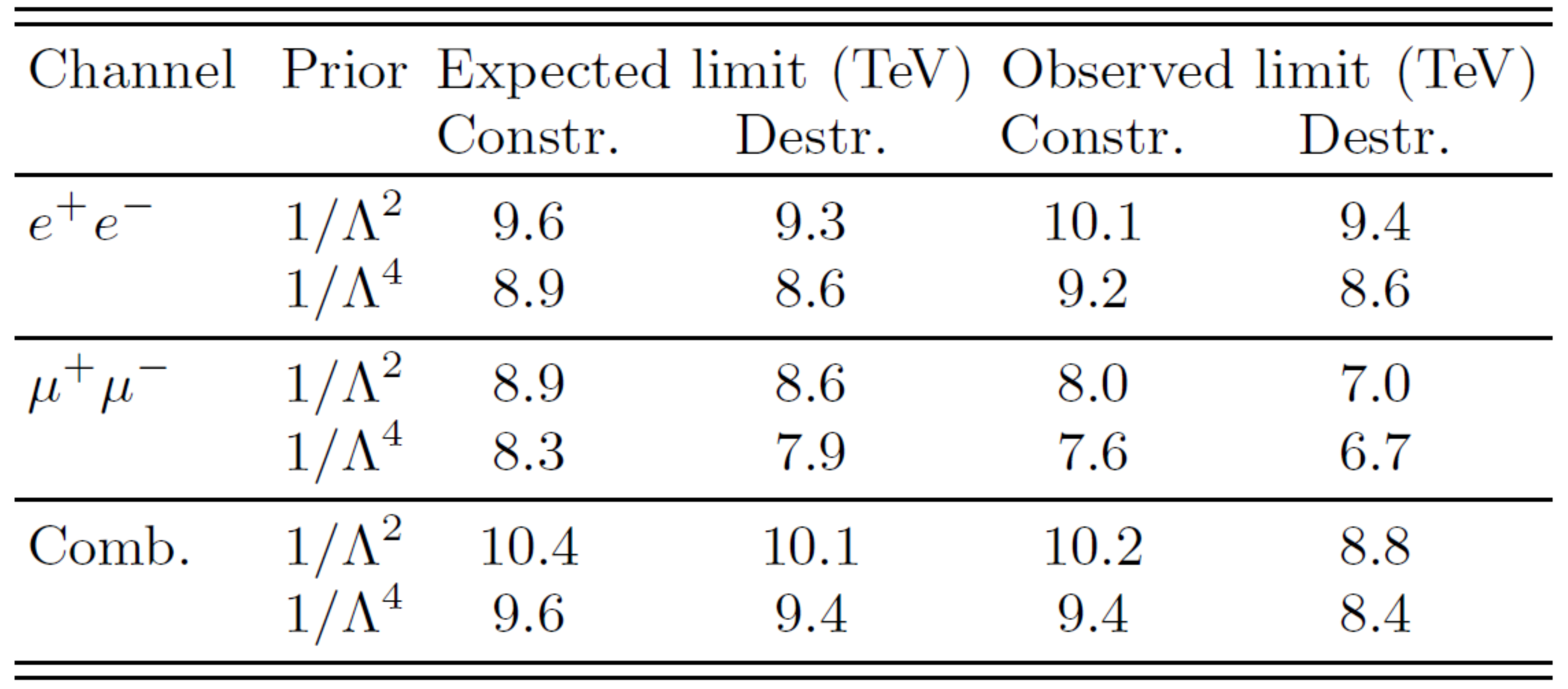} }
\label{tab:citable}       
\end{center}
\end{table}

\section{Conclusion}
While there is still no direct evidence of new phenomena at the LHC, the ATLAS Collaboration has set very stringent limits on many exotic models, analysing 1.03-1.94 fb$^{-1}$ of the LHC data. Some of these limits are the most stringent to date, or are novel models never tested by any other experiment.

%

\end{document}